# Does ChatGPT score research quality differently by gender?

Kayvan Kousha
Statistical Cybermetrics and Research Evaluation Group, Business School, University of Wolverhampton, UK. https://orcid.org/0000-0003-4827-971X

Mike Thelwall
School of Information, Journalism and Communication, University of Sheffield, UK. https://orcid.org/0000-0001-6065-205X; m.a.thelwall@sheffield.ac.uk

Large Language Models (LLMs) are being considered for research evaluation, raising concerns about the introduction of AI bias. This study investigates whether ChatGPT research quality scores differ by first-author gender using 89,744 journal articles from the UK Research Excellence Framework (REF) 2021. Author information was withheld from ChatGPT to avoid direct gender bias. Nevertheless, male first-authored papers had slightly higher ChatGPT scores in most Units of Assessment (UoAs), especially in health, science and engineering-related subjects, and this pattern was often stronger for ChatGPT than for REF scores, based on a departmental-level proxy. Rank-based ChatGPT gains relative to REF scores were also more favourable for male first-authored papers in most UoAs, although the differences were generally small. Gender differences were not evident for solo research in the social sciences, arts and humanities, however. The male-favouring pattern for first-authored research was not explained by gender differences in writing styles, at least as reflected in abstract complexity. Some ChatGPT–REF differences may also reflect the departmental averaging process used to generate the REF proxy scores. Average ChatGPT scores may differ by first-author gender indirectly through other factors, such as field, topic, method, journal context or authorship structure. Thus, this is an additional reason to be cautious with AI-based research evaluation.

**Keywords:** *Large language models; ChatGPT; research evaluation; REF2021; AI bias; gender; peer review*

## Introduction

Large language models (LLMs), including ChatGPT, are increasingly being evaluated as tools to support research evaluation (e.g., Thelwall, 2025, 2026). They would be especially useful for large-scale assessment systems such as the UK Research Excellence Framework (REF), where many outputs must be assessed across a wide range of fields (Sivertsen, 2017). However, using LLMs for research evaluation also creates important risks. Research quality is complex and involves judgements about originality, significance and rigour (Langfeldt et al., 2020). These judgements are difficult even for expert reviewers, and, lacking understanding, LLMs may rely on deep text pattern matching rather than genuine evaluation. Previous studies have found positive associations between ChatGPT scores and writing style (Kousha & Thelwall, 2026), publication year, country, length, or field (Thelwall & Kurt, 2025), suggesting that there is potential for AI bias in the scores.

Given ongoing concerns about gender biases in academia (Ceci et al., 2023; Eslen-Ziya & Yildirim, 2022) and research evaluation (Ray et al., 2024) as well as for AI (Hall & Ellis, 2023; O'Connor & Liu, 2024; Wan et al., 2023), it is important to know whether ChatGPT scores differ by author gender. Whilst direct gender bias can be eliminated or reduced by redacting author information before submitting an article for evaluation, author gender may be associated with other factors, such as field, topic, method, abstract writing style, or collaboration patterns (e.g., Lörz et al., 2011). For example, if ChatGPT gives higher scores to topics or methods that are more common among male-authored papers, then an apparent gender difference could emerge as a second-order effect. Similarly, if ChatGPT rewards male-oriented writing styles or methodological signals, then this would disadvantage other genders.

Recent evidence suggests that risks in LLM-assisted research evaluation are not only theoretical. For example, one study tested nine LLMs on 252 papers submitted to the International Conference on Learning Representations (ICLR) 2025 and changed only the author metadata in the review prompt. The study found small but model-dependent gender effects: Gemini and Llama tended to rate male-associated names higher, whereas ChatGPT-4o mini and Mistral tended to favour female-associated names (Vasu et al., 2026). Nevertheless, this evidence is limited because the study used a relatively small dataset from one machine learning conference. In economics peer review, one experiment changed author information for the same 330 papers. LLMs rated papers on a six-point review scale from "Definite Reject" to "Accept As Is". ChatGPT, Gemma and LLaMA rated papers attributed to top male economists higher than those attributed to top female economists or anonymised authorship, suggesting that gender and author reputation signals may affect LLM paper evaluations (Pataranutaporn et al., 2025). However, this result may partly reflect the use of famous economists rather than gender alone. There is also evidence that LLMs may have gender-related effects on reference selection. For instance, a controlled study using 660 articles from 22 research fields found that ChatGPT-4o selected more male-authored references in some settings (He, 2025).

There are substantial gender differences in academic research topics and even methods choices, which can affect any attempt to detect gender differences in academic-related tasks. In the UK, for example, female academics outnumber males 3 to 1 in Nursing and Allied Health Professions, but males outnumber females 6 to 1 in Electrical, Electronic, and Computer Engineering. In terms of UK publishing, female-authored Scopus articles outnumber male-authored in Veterinary Science by 2.3 to 1, whereas in mathematics the situation is the opposite by 3.7 to 1. At the most extreme level for narrow topics, females dominate Maternity and Midwifery by 19 to 1, compared to the reverse 12.5 to 1 for Geometry and Topology: a 237-fold difference. Disparities also occur within topics, such as with females being more likely to use qualitative methods or interviews in some fields (Thelwall et al., 2020). Since there are substantial field differences in citation rates, this can lead to female-authored research in any broad or narrow field attracting different average citation rates because topics more often published by female authors may be cited differently from topics more often published by male authors; a similar effect could potentially occur for AI given its apparent topic (Thelwall & Nunkoo, 2025; Thelwall, 2025) and field (Thelwall, 2026) biases in research quality scoring.

This study investigates whether ChatGPT research quality scores differ by first-author gender. The data consist of journal articles submitted to the UK Research Excellence Framework (REF) 2021. It compares female and male first-authored papers across REF Units of Assessment (UoAs: broad fields or clusters of similar fields) comparing ChatGPT scores to REF scores (using a proxy) and investigating linguistic complexity as a potential explanatory factor. The aim is to investigate whether the introduction of LLMs into research evaluation may tend to gender bias the results whatever the underlying cause, driven by the following research questions.

- **RQ1.** Do mean ChatGPT scores differ between female and male first-authored papers?
- **RQ2.** Does ChatGPT gain relative to REF differ between female and male first-authored papers?

# Methods

## *ChatGPT and REF scores*

The analysis used REF 2021 journal articles, ChatGPT scores, REF score proxies, first-author gender information, citation indicators and abstract readability indicators. The ChatGPT scores were taken from the REF2021 dataset from a previous large-scale study (Thelwall, 2026). This dataset contains ChatGPT research quality scores for REF2021 journal articles based on article titles and abstracts. The REF includes journal articles (and other outputs, not assessed here) from all fields and from all UK universities. The scoring was carried out through the OpenAI API using zero-shot prompts based on REF quality levels and descriptions. These were adapted from the REF2021 reviewer criteria, asking the model to assess originality, significance and rigour and to assign a score on the REF 1* to 4* quality scale. This scale ranges from nationally recognised quality at 1* to world-leading quality at 4*. Since the REF policy involves destroying the scores for individual articles, a proxy quality indicator was used instead. This is the average score of the articles in the submitting department(s) for the article. This can be calculated from departmental score profiles, which are public. These REF departmental average scores are not the same as individual article expert scores, hence the measure is an imperfect proxy. Its disadvantages are that (a) it dampens (reduces) the strengths of underlying correlations with other variables due to the replacement of individual scores with averages and (b) it is susceptible to departmental level biases due to specialisms if ChatGPT has a bias towards a type of research that is disproportionately produced by one department. It is nevertheless the largest-scale available research quality indicator and has been used in many previous studies (e.g., Kousha & Thelwall, 2026; Thelwall, 2026).

For the gender analysis, ChatGPT scores and REF proxy scores were first compared separately between female and male first-authored papers, generating a dataset of 89,744 REF 2021 journal article submissions with first-author gender information. Some articles had been evaluated by more than one REF UoA. Appendix A reports the female and male first-author counts by UoA.

For the rank-gain analysis, ChatGPT and REF proxy scores were converted into percentile ranks separately within each UoA. ChatGPT rank gain was then calculated as the ChatGPT percentile rank minus the REF percentile rank. Positive values indicate that a paper was ranked more favourably by ChatGPT than by the REF proxy, while negative values indicate the opposite. This rank-based approach was used because ChatGPT scores and REF proxy scores have different score distributions and should not be directly compared as raw score differences.

## *First-author gender identification*

Although most research is produced by teams of people, often authors with different genders, the analysis associates the gender of the first author with the gender of the article. This is because the first author is both the most visible and consistently one of the main contributors to work (Larivière et al., 2016), although corresponding or last authors may better indicate research leadership in some fields.

For example, corresponding authors are usually first authors, but this is less common in Medicine, Natural Sciences and Engineering (Chinchilla-Rodríguez et al., 2024). To reduce this ambiguity, a further analysis was also conducted for solo-authored papers, where the first author is necessarily the only author (see Discussion).

First-author gender was gathered from first names using the available gender-identification fields in the dataset. A large list of gendered first names from the UK and USA (Thelwall et al., 2020) was used for this. Each name in the list is associated at least 95% of the time with either males or females, excluding more gender-neutral names like Sam and Pat. When an author had a name in this list, they were assumed to have the associated gender. Otherwise, their publication was removed from the data analysed.

## *ChatGPT scores compared to REF scores*

Comparing ChatGPT with REF scores is not simple. Although they both fall within the same range of 1* to 4*, they do not necessarily have the same means and variances. Previous studies with ChatGPT quality scores (e.g., Thelwall, 2026) have found it to give different mean scores to human reviewers and to have a much smaller variance. On this basis, the relevant ChatGPT information is not its (inaccurate) scores, but the rank order of the articles based on their ChatGPT scores. Thus, the appropriate way to identify any gender effect of a change from REF scores to ChatGPT scores is to see if the average rank (rather than average score) of papers changes by gender.

Articles in each UoA were therefore ranked twice: once by ChatGPT score and once by REF score. The REF rank was then subtracted from the ChatGPT rank and a Mann-Whitney U test used to see if the rank differences were statistically significant between males and females.

For the gender analysis, ChatGPT scores and (departmental average) REF expert scores were first compared separately between female and male first-authored papers. For the rank-gain analysis, ChatGPT and REF scores were converted into percentile ranks separately within each UoA. The ChatGPT rank gain was then calculated as the ChatGPT percentile rank minus the REF percentile rank. This rank-based approach was used because ChatGPT scores and REF proxy scores have different score distributions and should not be directly compared as raw score differences.

## Results

### *Mean ChatGPT and REF scores differences by first-author gender*

Figure 1 illustrates female minus male differences in mean ChatGPT and REF proxy scores across the 34 REF Units of Assessment. Negative values mean that male first-authored papers had higher mean scores, whereas positive values mean that female first-authored papers had higher mean scores. In most UoAs, ChatGPT and REF scores had the same direction of female minus male score difference. In the context of historical female disadvantages in academia, it is discouraging that there is an overall male advantage pattern in both sets of research quality scores. The average female minus male difference was generally larger for ChatGPT than for REF especially across several STEM-related UoAs, but this difference could be due to differing score variances so the magnitude of the two differences should not be directly compared.

Although there are some fields where the direction of the gender advantage varies between REF scores and ChatGPT scores, in all cases at least one of the two 95% confidence intervals contains 0 (e.g., Business, Sport), so the results do not give clear evidence that ChatGPT introduces any direction of gender bias relative to REF scores.

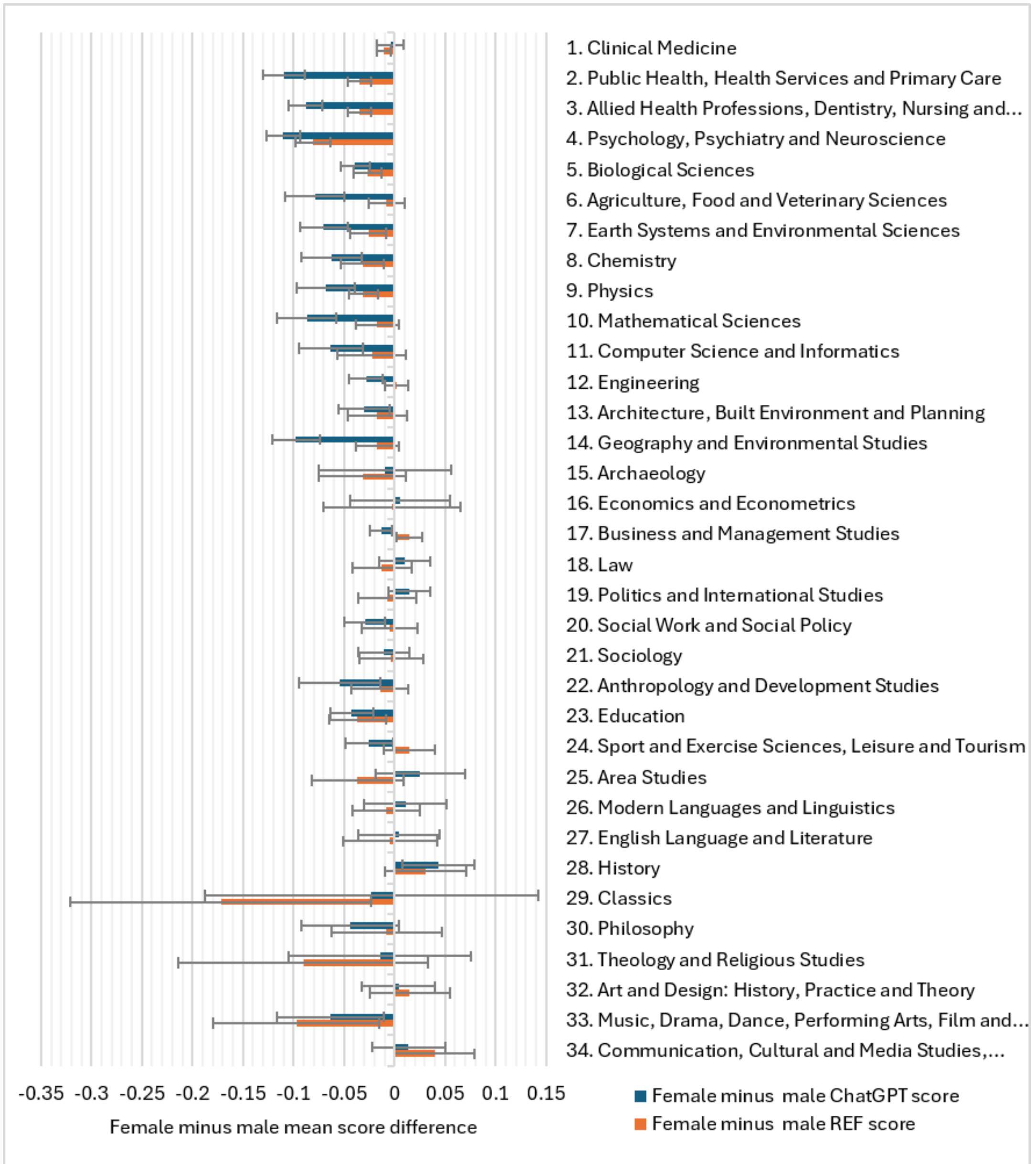

**Figure 1. Female minus male mean score differences for ChatGPT and REF proxy scores by UoA.** Positive values indicate higher mean scores for female first-authored papers; negative values indicate higher mean scores for male first-authored papers.

## *Gender differences in ChatGPT rank gain relative to REF*

Figure 2 compares female minus male mean ranks for ChatGPT rank gain relative to the REF average departmental score across the 34 UoAs. This analysis first ranked papers separately by ChatGPT score and REF score within each UoA and then assessed whether female or male first-authored papers improved their rank most under ChatGPT relative to the REF rank. Rank orders are fair to compare between the different scores because they are not influenced by score means and variances. Negative values indicate higher ChatGPT rank gains for male first-authored papers, while positive values indicate higher rank gains for female first-authored papers.

Male first-authored papers had higher mean ranks for ChatGPT rank gain in 23 of the 34 UoAs, while female first-authored papers had higher mean ranks in the remaining 11 UoAs. Although the direction was more often male-favourable for ChatGPT, many UoA-level differences were small and not statistically significant. The Mann–Whitney U tests found statistically significant differences in 13 UoAs. Eleven of these were male-favourable, while only Clinical Medicine and Area Studies were female-favourable. The results therefore suggest that ChatGPT rank gain relative to REF was more often favourable for male first-authored papers, although in many UoAs differences were small and not statistically significant.

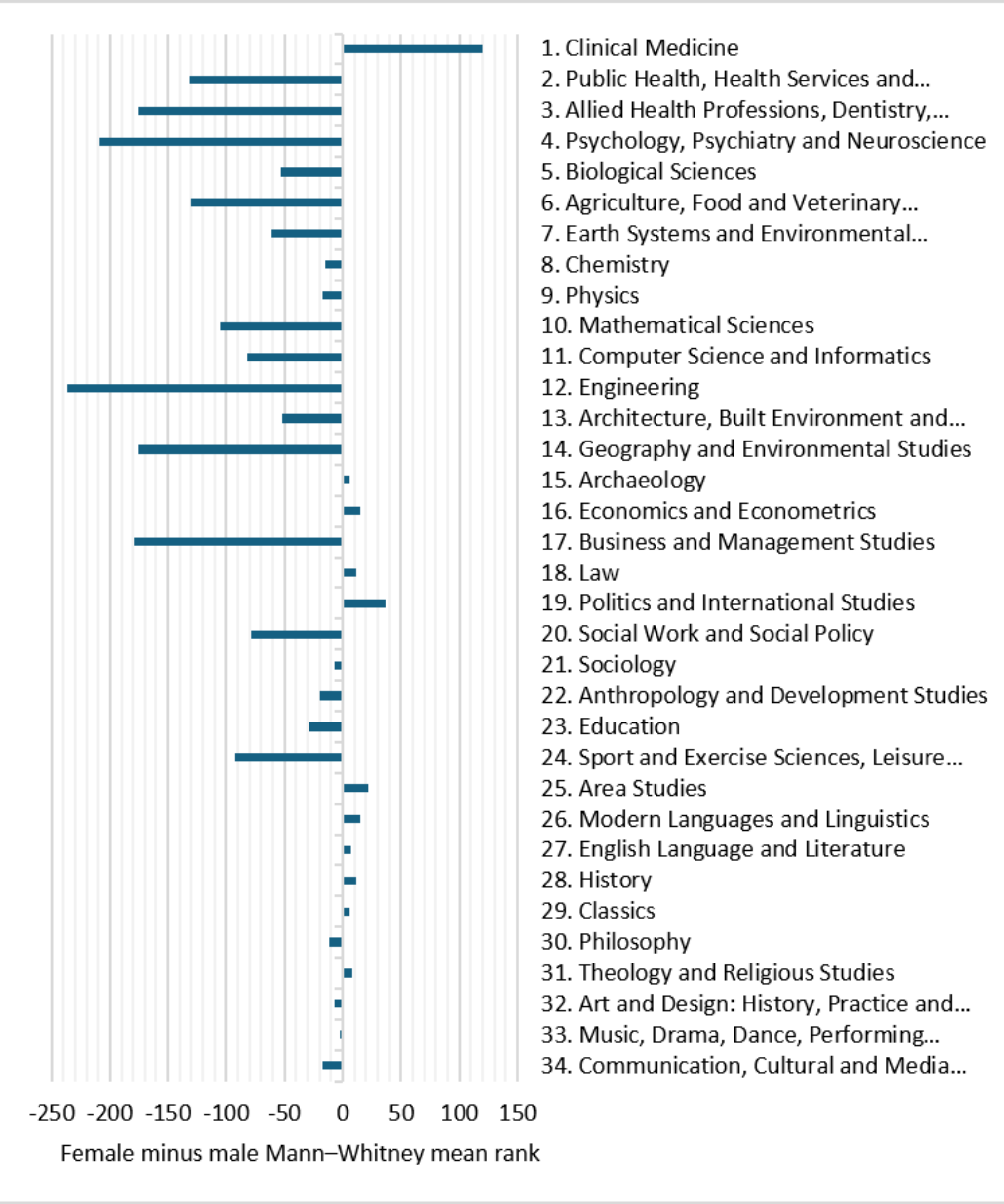

**Figure 2. Gender differences in ChatGPT rank gain relative to REF across 34 UoAs.** Values show female minus male Mann–Whitney mean ranks (i.e., ChatGPT rank gain relative to REF scores). Positive values indicate ChatGPT female-favourable rank gain; negative values indicate ChatGPT male-favourable rank gain.

# Discussion

The main results suggest that male first-authored papers often had slightly higher ChatGPT scores and more favourable ChatGPT scores relative to REF departmental average peer review scores, especially in several STEM-related UoAs. The following discussion examines whether this pattern may partly reflect other factors associated with first-author gender, including citation impact, abstract complexity patterns.

## *Citation impact differences by first-author gender*

Figure 3 shows female minus male differences in NLCS 2024 across the 34 REF Units of Assessment. Positive values indicate higher normalised citation impact for female first-authored papers, whereas negative values show higher normalised citation impact for male first-authored papers. The female minus male NLCS 2024 difference was negative in 21 of the 34 UoAs and positive in the remaining 13 UoAs. The average UoA-level difference was small and negative, about -0.023, with a median of about -0.019. This suggests that male first-authored papers had slightly higher normalised citation impact in more UoAs, although the pattern was not uniform across all fields. Panel A UoAs had negative differences. Most Panel B UoAs were also negative, except Computer Science and Informatics, where the difference was slightly positive. This suggests that in health, life sciences and much of STEM, male first-authored papers tended to have higher normalised citation impacts.

Panels C and D were more mixed, perhaps because of smaller sample sizes and less importance for citations. In Panel C, several UoAs had positive female minus male citation-impact differences, including Economics and Econometrics, Law, Social Work and Social Policy, and Sociology. In Panel D, the pattern was also mixed, with relatively positive differences in Theology and Religious Studies and Music, Drama, Dance, Performing Arts, Film and Screen Studies, but a large negative difference in Philosophy.

This result is useful because it partly overlaps with the earlier ChatGPT gain findings (Figure 2), where male first-authored papers often had a favourable ChatGPT rank gain relative to REF, especially in most STEM-related UoAs. However, the NLCS pattern does not fully match the ChatGPT pattern. Several social science and humanities UoAs had higher citation impact for female first-authored papers, while the earlier ChatGPT results were often still male-favourable or mixed. This suggests that citation impact gender differences may mimic part of the ChatGPT gender pattern, but not all of it.

This result is inconsistent with a study using individual REF peer-review scores and converted NLCS into REF-like bibliometric scores (Thelwall et al., 2023), finding a small female gain from bibliometrics. However, the current raw NLCS analysis shows slightly higher citation impact for male first-authored papers in more UoAs. This could be because the previous study measured bibliometric gain relative to individual REF scores and used individual article level REF scores, whereas Figure 3 measures raw normalised citation impact using a departmental REF score proxy. Thus, the results in the current paper could be an artefact of the departmental proxy since aggregating results at the departmental level reversed the direction of the citation gender inequality.

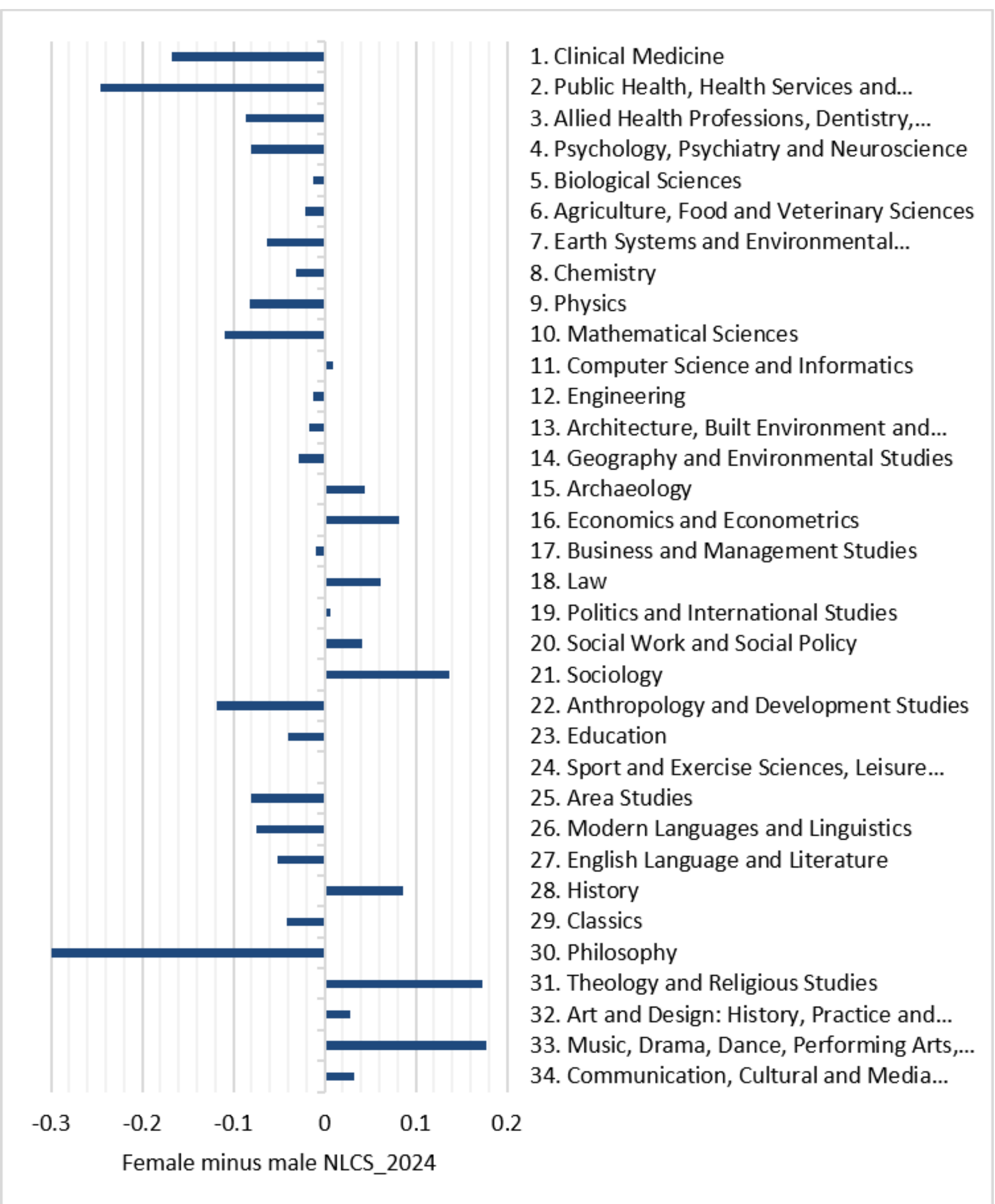

**Figure 3. Female minus male differences in NLCS 2024 by REF Unit of Assessment.** Positive values indicate higher citation impact for female first-authored papers. Negative values indicate higher citation impact for male first-authored papers.

## *Abstract complexity by first-author gender*

Since there are gender differences in writing styles (Nugroho & Suseno, 2025; Rubin & Greene, 1992), we checked whether first-author female/male score differences could be related to abstract writing style. The Flesch–Kincaid Grade (Mailloux et al., 1995) was used as a measure of abstract complexity. Higher Flesch–Kincaid Grade values indicate more complex or less readable abstracts because the measure is based on sentence length and syllables per word. Abstract complexity data were matched from a previous REF2021 abstract readability study (Kousha & Thelwall, 2026). The study calculated various readability and textual indicators for REF2021 journal article abstracts using a Python script based on the *textstat library* (https://pypi.org/project/textstat/). The current study used Flesch–Kincaid Grade as the main abstract complexity indicator. These data were used to check whether gender differences in ChatGPT scores could partly reflect gender differences in abstract complexity. Figure 4 compares the female minus male differences in Flesch–Kincaid Grade across the 34 REF Units of Assessment. Positive values mean that female first-authored abstracts had higher average Flesch–Kincaid Grade scores, whereas negative values mean that male first-authored abstracts had higher average scores. The female minus male difference was positive in 20 of the 34 UoAs and negative in the remaining 14 UoAs. The pattern was mixed in Panel A and mostly negative in Panel B, but it was mostly positive in Panels C and D. This suggests that female first-authored abstracts were not generally simpler or more readable than male first-authored abstracts. In several arts, humanities and social science UoAs, female first-authored abstracts had higher average Flesch–Kincaid Grade scores. This result is important because it does not clearly explain the male first-author pattern in Figures 1 and 2 are related to writing style of female first-authors. If ChatGPT tended to give higher scores to more complex abstracts (Kousha & Thelwall, 2026), then higher Flesch–Kincaid Grade scores for female first-authored abstracts would be expected to favour female first-authored papers in these UoAs.

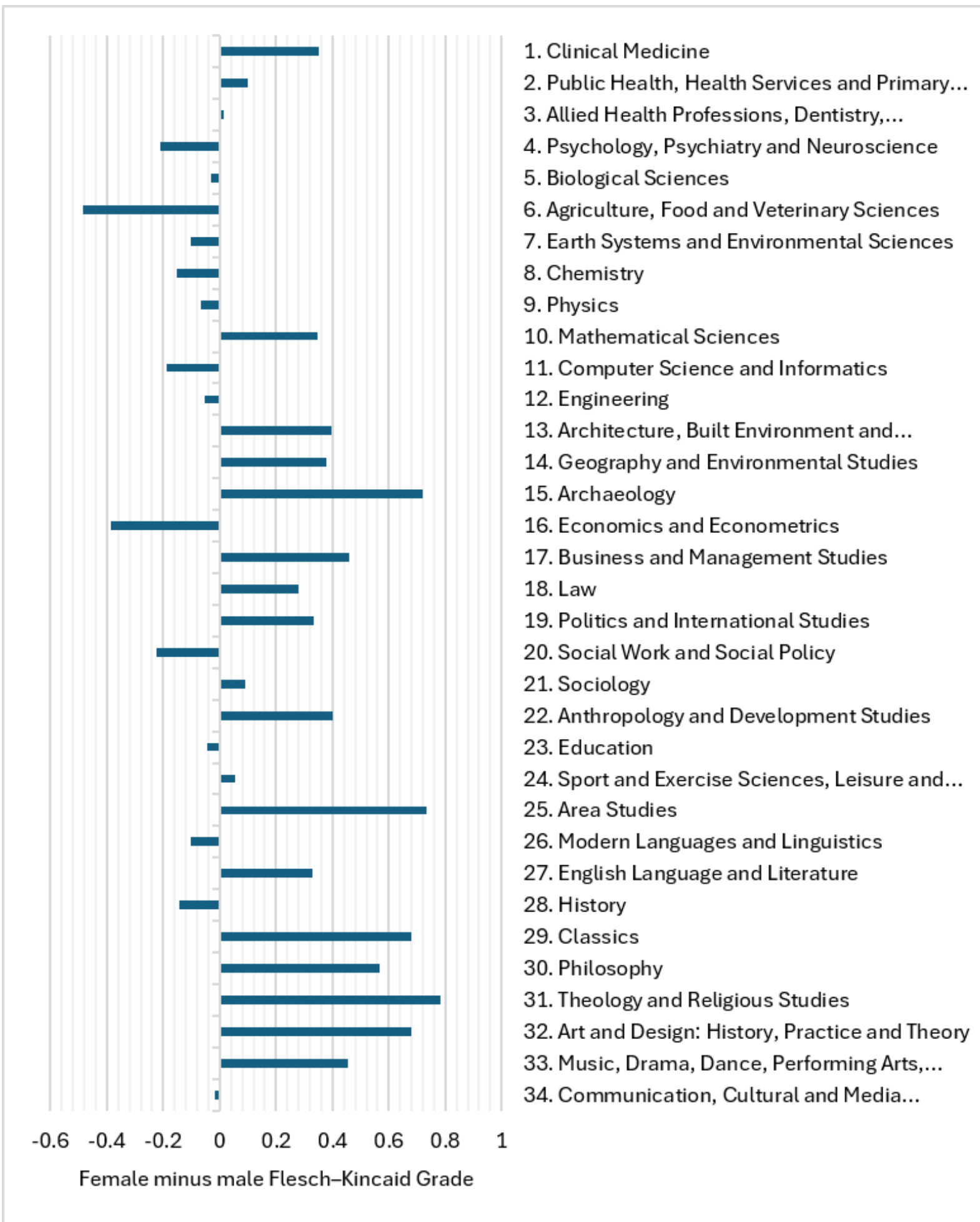

**Figure 4. Female minus male differences in Flesch–Kincaid Grade by REF Unit of Assessment.** Positive values indicate higher average abstract complexity for female first-authored papers; negative values indicate higher average abstract complexity for male first-authored papers.

### *Other possible reasons for first-author gender differences*

Previous evidence has suggested that first-author gender differences in research assessment scores can reflect gender differences in fields, topics and methods rather than gender bias (Thelwall et al., 2020). This is relevant because previous ChatGPT studies found that ChatGPT tends to score some methods and topics differently, including higher scores for theory, statistics, experiments and algorithms, and lower scores for surveys or convenience sampling (Thelwall, 2025; Thelwall & Nunkoo, 2025). Hence, Figure 4 suggests that the male first-author ChatGPT advantage may partly reflect differences in citation impact, topic choice, methods or disciplinary publishing patterns.

### *Solo-authored papers*

To partly assess whether the first-author analysis was affected by team authorship, the ChatGPT rank-gain analysis in Figure 2 was repeated for solo-authored papers (Figure 5). This reduces ambiguity in the gender indicator because, in solo-authored papers, the first author is also the only author, whereas in multi-authored papers authors with different genders may make substantial contributions (e.g., the last author in some fields). Solo-authored research may also be a different type to team research (e.g., less complex, more theoretical), however, so a solo authorship analysis is not exactly equivalent to the previous all-authorship analysis. The solo-author analysis was checked for Panels C and D. Panels A and B had relatively small numbers of solo-authored papers, especially for females and hence they were not used as strong evidence. Appendix A reports the solo-author counts by UoA.

In Panel C, the solo-author results did not indicate a clear male or female-favourable ChatGPT rank-gain pattern. However, the Panel D solo-author analysis gave a different pattern from the first-author analysis. In the first-author Panel D analysis, six of the ten UoAs were female-favourable and four were male-favourable (Figure 5). In contrast, among solo-authored papers, nine of the ten Panel D UoAs were female-favourable, with only English Language and Literature having a (very small) male-favourable difference. Several UoAs that were male-favourable in the first-author analysis, including Philosophy, Art and Design, Music and Drama, and Communication, Cultural and Media Studies, became female-favourable in the solo-author analysis. Nevertheless, the analysis should be interpreted cautiously because none of the Panel D solo-author UoA-level differences were statistically significant and Mann–Whitney mean-rank differences should be interpreted mainly in terms of direction rather than magnitude because they are affected by sample size.

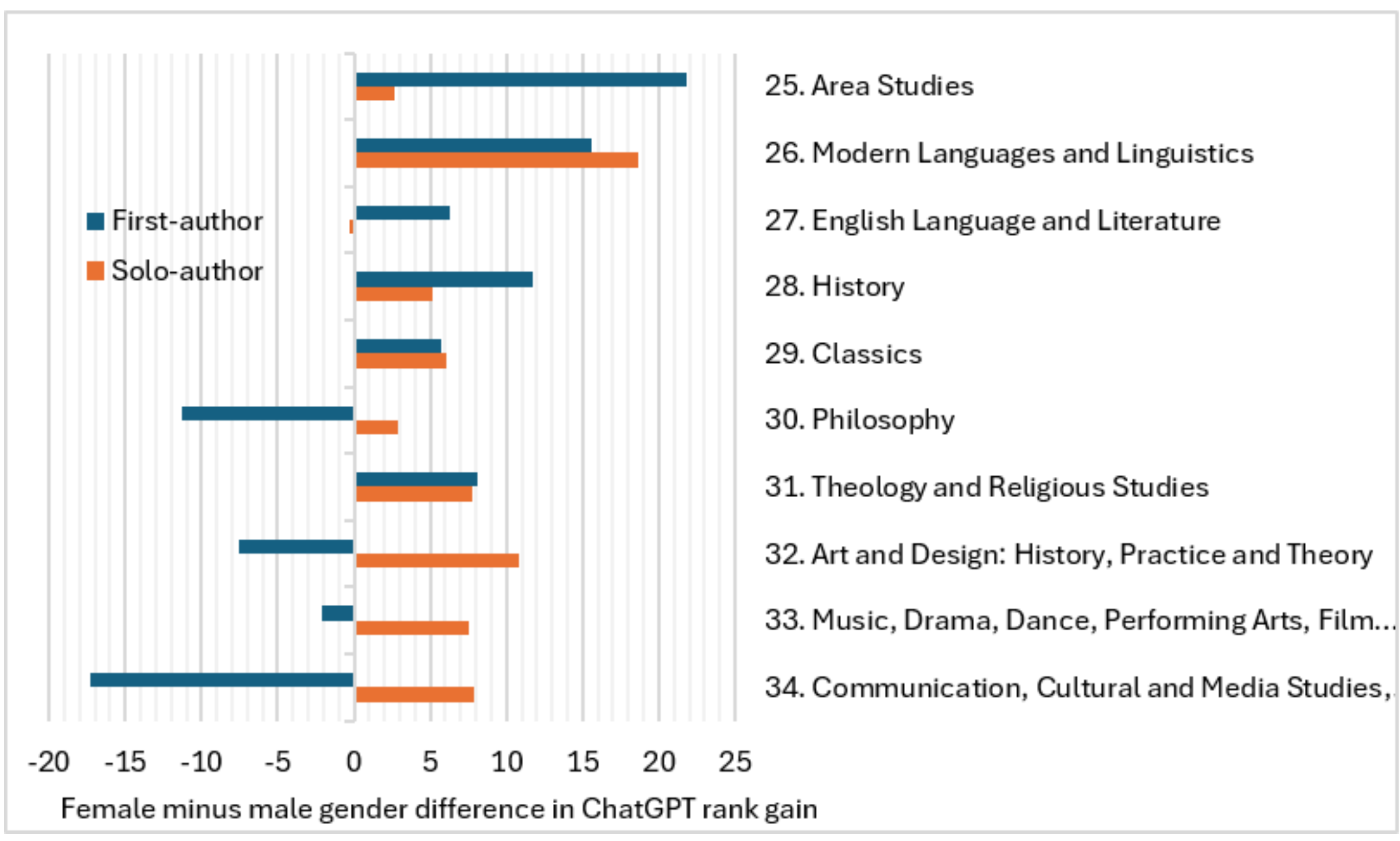

**Figure 5. First-author and solo-author gender differences in ChatGPT rank gain relative to REF in Panel D.**

We also analysed stylistic features of abstracts for solo-authored papers to assess if abstract complexity differed between female and male at UoA level within Panels C and D (Figure 6). Positive values indicate higher average abstract complexity for female solo-authored papers, whereas negative values indicate higher average abstract complexity for male solo-authored papers. Across the 22 UoAs in Panels C and D, the female minus male Flesch–Kincaid difference was positive in 15 UoAs and negative in the remaining 7 UoAs. This suggests that female solo-authored abstracts were usually more complex or less readable than male solo-authored abstracts. Hence, the result does not support the explanation that male solo-authored papers received more favourable ChatGPT scores because their abstracts were more complex.

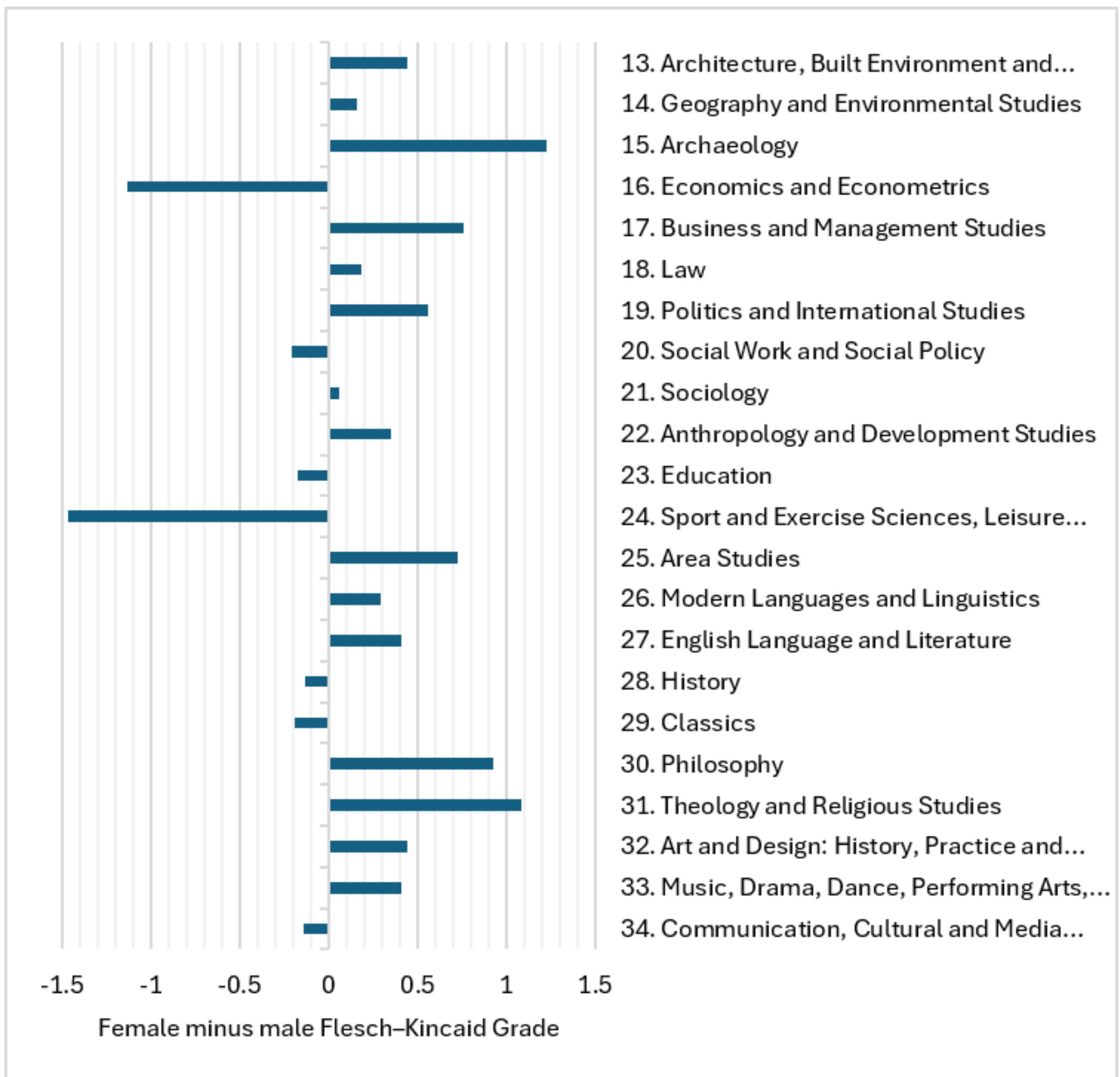

**Figure 6. Female minus male differences in Flesch–Kincaid Grade for solo-authored papers in Panels C and D.** Positive values indicate higher average abstract complexity for female solo-authored papers; negative values indicate higher average abstract complexity for male solo-authored papers.

## Limitations

This study has several limitations. First author gender was estimated from first names using the existing name list from the UK and USA. Authors with unknown, ambiguous or gender-neutral names were excluded, and many unknown first names were initials only as extracted from the Scopus database, so gender could not be identified. First-author gender is also an imperfect indicator for team-authored papers because many papers have several authors and author roles vary by field. In some subjects, corresponding or last authors may better indicate research leadership than first authors (Chinchilla-Rodríguez et al., 2024). The solo-author analysis partly reduces this problem, but solo-authored papers are less common in STEM fields (see also Appendix A) and may differ from team-authored research in terms of type of research, methods and scopes which may potentially influence the research quality scores. The study also used departmental average REF scores as a proxy because individual article-level REF scores were destroyed after REF2021. This allows a large-scale practical comparison, but it may hide article-level variation and introduce departmental effects.

## Conclusions

The first-author analysis showed a male-favourable pattern in ChatGPT scores, especially in health, science and engineering-related UoAs. However, the solo-author analysis for Panels C and D did not suggest that ChatGPT systematically favoured either female or male solo-authored papers. Abstract complexity was not found to be a likely cause of the difference. The interpretation of the differences between ChatGPT and REF scores is complicated by the fact that the REF score proxy used found a different direction of gender difference for citations compared to a previous study that used direct REF scores (now destroyed), which undermines the strength of the evidence in the current paper in terms of differences between ChatGPT and REF scores.

In answer to the first research question, mean ChatGPT and REF scores differed by first-author gender in several UoAs (Figure 1). In most UoAs, male first-authored papers had slightly higher average ChatGPT scores and this pattern was often stronger for ChatGPT than for the REF score proxy. However, the differences were generally small.

In answer to the second research question, ChatGPT gain relative to REF also differed by first-author gender (Figure 2). Male first-authored papers had a more favourable ChatGPT gain in most UoAs, especially in health, science and engineering-related areas. However, the solo-author checks did not reproduce a clear male-favourable pattern in Panels C and D.

The results suggest that the male-favourable ChatGPT pattern is unlikely to be explained by abstract complexity alone. It may reflect other factors linked to first author gender such as topic, method, field, journal context or authorship structure. For example, if ChatGPT tends to give higher scores to papers using statistical methods or experiments and these papers are more often male first-authored in some UoAs, then a male-favourable pattern could appear even when author names are not included in ChatGPT feeds (using only title and abstracts). In contrast, if female first-authored papers are more common in qualitative research and ChatGPT scores these approaches less highly, then this could also create gender-related score differences. Thus, this would be an indirect gender effect rather than direct gender bias based on authors' names.

In conclusion, the results show male-favouring ChatGPT score associations in more UoAs than the reverse, although they do not give a likely explanation for it. Thus, future users of LLMs for research quality scoring should check for gender-related patterns, especially where scores may be influenced by field, topic, method, journal context or authorship structure.

## References

Ceci, S. J., Kahn, S., & Williams, W. M. (2023). Exploring gender bias in six key domains of academic science: An adversarial collaboration. Psychological Science in the Public Interest, 24(1), 15-73.

Chinchilla-Rodríguez, Z., Costas, R., Robinson-García, N., & Larivière, V. (2024). Examining the quality of the corresponding authorship field in Web of Science and Scopus. Quantitative Science Studies, 5(1), 76-97.

Eslen-Ziya, H., & Yildirim, T. M. (2022). Perceptions of gendered-challenges in academia: How women academics see gender hierarchies as barriers to achievement. *Gender, Work & Organization*, *29*(1), 301-308.

Hall, P., & Ellis, D. (2023). A systematic review of socio-technical gender bias in AI algorithms. Online Information Review, 47(7), 1264-1279.

He, J. (2025). Who gets cited? Gender- and majority-bias in LLM-driven reference selection. arXiv preprint arXiv:2508.02740.

Kousha, K., & Thelwall, M. (2026). Which stylistic features fool ChatGPT research evaluations? Scientometrics. https://doi.org/10.1007/s11192-026-05756-1

Langfeldt, L., Nedeva, M., Sörlin, S., & Thomas, D. A. (2020). Co-existing notions of research quality: A framework to study context-specific understandings of good research. *Minerva*, 115-137.

Larivière, V., Desrochers, N., Macaluso, B., Mongeon, P., Paul-Hus, A., & Sugimoto, C. R. (2016). Contributorship and division of labor in knowledge production. *Social studies of science*, *46*(3), 417-435.

Lörz, M., Schindler, S., & Walter, J. G. (2011). Gender inequalities in higher education: Extent, development and mechanisms of gender differences in enrolment and field of study choice. *Irish Educational Studies*, *30*(2), 179-198.

Mailloux, S. L., Johnson, M. E., Fisher, D. G., & Pettibone, T. J. (1995). How reliable is computerized assessment of readability? *Computers in Nursing*, *13*, 221-221.

Nugroho, M., & Suseno, I. G. (2025). Gender and language: Analyzing communication styles in argumentative writing. *Journal of Communication and Public Relations*, *4*(1), 102-117.

O'Connor, S., & Liu, H. (2024). Gender bias perpetuation and mitigation in AI technologies: challenges and opportunities. *AI & SOCIETY*, *39*(4), 2045-2057.

Pataranutaporn, P., Powdthavee, N., Achiwaranguprok, C., & Maes, P. (2025). Can AI Solve the Peer Review Crisis? A Large Scale Cross Model Experiment of LLMs' Performance and Biases in Evaluating over 1000 Economics Papers. arXiv preprint arXiv:2502.00070.

Ray, K. S., Zurn, P., Dworkin, J. D., Bassett, D. S., & Resnik, D. B. (2024). Citation bias, diversity, and ethics. *Accountability in Research*, *31*(2), 158-172.

Rubin, D. L., & Greene, K. (1992). Gender-typical style in written language. Research in the Teaching of English, 26(1), 7-40.

Sivertsen, G. (2017). Unique, but still best practice? The Research Excellence Framework (REF) from an international perspective. *Palgrave Communications*, *3*(1), 17078.

Thelwall, M. (2025). Can ChatGPT replace citations for quality evaluation of academic articles and journals? Empirical evidence from library and information science. Journal of Documentation, 81(4), 1078-1094.

Thelwall, M. (2026). In which fields do ChatGPT scores align more closely with research quality than do citation rates? Journal of Data and Information Science.

Thelwall, M., & Nunkoo, R. (2025). Evaluating the quality of tourism research using ChatGPT. Current Issues in Tourism, 1-25.

Thelwall, M., Abdoli, M., Lebiedziewicz, A., & Bailey, C. (2020). Gender disparities in UK research publishing: Differences between fields, methods and topics. Profesional de la Información, 29(4),1699-2407.

Thelwall, M., Kousha, K., Stuart, E., Makita, M., Abdoli, M., Wilson, P., & Levitt, J. (2023). Do bibliometrics introduce gender, institutional or interdisciplinary biases into research evaluations? Research Policy, 52(8), 104829.

Thelwall, M., & Kurt, Z. (2025). Research evaluation with ChatGPT: Is it age, country, length, or field biased? Scientometrics, 130(10), 5323-5343.

Vasu, S. S. M., Sheth, I., Wang, H. P., Binkyte, R., & Fritz, M. (2026). Justice in judgment: Unveiling (hidden) bias in llm-assisted peer reviews. In Findings of the Association for Computational Linguistics: ACL 2026 (pp. 307-330).

Wan, Y., Pu, G., Sun, J., Garimella, A., Chang, K. W., & Peng, N. (2023, December). “Kelly is a warm person, Joseph is a role model”: Gender biases in LLM-generated reference letters. In *Findings of the Association for Computational Linguistics: EMNLP 2023* (pp. 3730-3748).

**Appendix A. Number of solo-authored and first-authored journal articles by gender across REF UoAs.**

| UoA | Solo female-author | Solo male-author | Female first-author | Male first-author |
|---|---|---|---|---|
| 1. Clinical Medicine | 2 | 8 | 3777 | 4847 |
| 2. Public Health, Health Services and Primary Care | 9 | 11 | 1790 | 1868 |
| 3. Allied Health Professions, Dentistry, Nursing and Pharmacy | 41 | 17 | 4024 | 3617 |
| 4. Psychology, Psychiatry and Neuroscience | 20 | 63 | 3541 | 3545 |
| 5. Biological Sciences | 1 | 12 | 2268 | 3131 |
| 6. Agriculture, Food and Veterinary Sciences | 3 | 9 | 1225 | 1284 |
| 7. Earth Systems and Environmental Sciences | 17 | 38 | 1095 | 2059 |
| 8. Chemistry | 0 | 4 | 703 | 1471 |
| 9. Physics | 4 | 21 | 584 | 2573 |
| 10. Mathematical Sciences | 20 | 188 | 420 | 2139 |
| 11. Computer Science and Informatics | 11 | 71 | 546 | 2391 |
| 12. Engineering | 6 | 148 | 2103 | 7620 |
| 13. Architecture, Built Environment and Planning | 103 | 166 | 602 | 1337 |
| 14. Geography and Environmental Studies | 200 | 236 | 1041 | 1790 |
| 15. Archaeology | 27 | 34 | 183 | 250 |
| 16. Economics and Econometrics | 16 | 58 | 128 | 490 |
| 17. Business and Management Studies | 255 | 434 | 3018 | 5330 |
| 18. Law | 409 | 452 | 667 | 718 |
| 19. Politics and International Studies | 331 | 515 | 671 | 1186 |
| 20. Social Work and Social Policy | 358 | 313 | 1486 | 1104 |
| 21. Sociology | 238 | 226 | 573 | 552 |
| 22. Anthropology and Development Studies | 133 | 141 | 308 | 326 |
| 23. Education | 356 | 314 | 1477 | 1136 |
| 24. Sport and Exercise Sciences, Leisure and Tourism | 25 | 60 | 743 | 1621 |
| 25. Area Studies | 85 | 102 | 154 | 188 |
| 26. Modern Languages and Linguistics | 193 | 132 | 388 | 253 |
| 27. English Language and Literature | 194 | 142 | 258 | 194 |
| 28. History | 209 | 327 | 248 | 391 |
| 29. Classics | 21 | 11 | 26 | 22 |
| 30. Philosophy | 76 | 217 | 97 | 289 |
| 31. Theology and Religious Studies | 31 | 64 | 39 | 72 |
| 32. Art and Design: History, Practice and Theory | 139 | 87 | 368 | 318 |
| 33. Music, Drama, Dance, Performing Arts, Film and Screen Studies | 104 | 146 | 158 | 235 |
| 34. Communication, Cultural and Media Studies, Library and Information Management | 180 | 159 | 353 | 335 |
| Total | 3817 | 4926 | 35062 | 54682 |